\documentclass[%
 reprint,
 superscriptaddress,
 amsmath,amssymb,
 aps,
]{revtex4-2}

\usepackage{graphicx, ragged2e, subcaption}
\usepackage{dcolumn}
\usepackage{bm}

\begin{document}

\preprint{APS/123-QED}



\title{Forecasting infectious diseases in Brazilian cities: integrating socio-economic and geographic data from related cities through a machine learning approach}


\author{L. Lober}
    \email{luiza.lober@usp.br}
    \affiliation{%
     Departamento de Matemática Aplicada e Estatística, Instituto de Ciências Matemáticas e de Computação, Universidade de São Paulo—Campus de São Carlos, Caixa Postal 668, 13560-970 São Carlos, São Paulo, Brazil
    }%

\author{K. Oliveira Roster}%
    \affiliation{%
     Harvard T. H. Chan School of Public Health, Boston, MA.
    }%
    
\author{F. A. Rodrigues}
    \affiliation{%
     Departamento de Matemática Aplicada e Estatística, Instituto de Ciências Matemáticas e de Computação, Universidade de São Paulo—Campus de São Carlos, Caixa Postal 668, 13560-970 São Carlos, São Paulo, Brazil
    }%

\date{\today}


\begin{abstract}       

    Supervised machine learning models and public surveillance data have been employed for infectious disease forecasting in many settings. These models leverage various data sources capturing drivers of disease spread, such as climate conditions or human behavior. However, few models have incorporated the organizational structure of different geographic locations for forecasting. Traveling waves of seasonal outbreaks have been reported for dengue, influenza, and other infectious diseases, and many of the drivers of infectious disease dynamics may be shared across different cities, either due to their geographic or socioeconomic proximity. In this study, we developed a machine learning model to predict case counts of four infectious diseases across Brazilian cities one week ahead by incorporating information from related cities. We compared selecting related cities using both geographic distance and GDP per capita. Incorporating information from geographically proximate cities improved predictive performance for two of the four diseases, specifically COVID-19 and Zika. We also discuss the impact on forecasts in the presence of anomalous contagion patterns and the limitations of the proposed methodology.

\end{abstract}


\maketitle


\section{Introduction}



Data driven models are increasingly used for disease forecasting given the availability of public health records and advances in machine learning algorithms and their applications to epidemiology \cite{Zhao2020Sep, Rahimi2023Nov}, with Brazil's \textit{DataSUS} \cite{Datasus_site} reporting over 50 unique diseases and health conditions being actively monitored, including multiple endemic diseases and also COVID-19, which had widespread repercussions in the country and beyond in recent years. 

When employing theoretical procedures, one has the possibility of understanding the dynamics and properties of the infectious agents through compartmental models, such as done analysing the different COVID-19 variants \cite{Dutta2022Jun} and Dengue's serotypes \cite{Andraud2012Nov}, with works also considering the co-circulation of diseases transmitted by the \textit{Aedes} mosquito \cite{Hirata2023Aug}. However, the main advantage of current supervised machine learning approaches over those approaches lies in their ability to directly infer the relationship between features of interest without the necessary knowledge of the epidemic's dynamics to arrive at reasonable predictions \cite{Roster2022Sep}. 

In terms of computational costs and explainability, decision trees ensembles, such as the Random Forests or the XGBoost algorithms, often proves to be a reliable method, specially when the neural networks' requirement of data availability is in the tens of thousands of measurements, which could also bias these models towards over-fitting.
 

A range of data sources has been used for forecasting, including active and passive surveillance records (cases, hospitalizations, serosurveys) \cite{Ebi2016Nov}, meteorological and environmental conditions \cite{Xu2020Mar}, human behavior and internet searches \cite{Moran2016Dec}. The importance of different data sources varies among infectious diseases.  

As a vector-borne disease, dengue cases depend on vector suitability conditions and human behavior, such as socioeconomic factors affecting vector suitability, mobility, and susceptibility to infection\cite{Shepard2011Feb}, with Brazil currently accounting for the most cases reported in Latin America, with an increase of 13$\%$ in these registers when compared to the past year \cite{WHO_Dengue_2023}. 

COVID-19 on the other hand is a respiratory condition and thus cases are more closely correlated with measures of human interactions, such as crowding in public spaces and traveling, with early lock-downs, mask and latter vaccination adoption also had significant implications to its development, with the rate, delays and general adoption of those methods varying significantly between countries \cite{Basak2022Feb, An2021Nov}. 

Concurrently, other endemic infectious diseases also affect the country, such as Zika virus, which was shown to be linked to Dengue and Chikungunya \cite{Pessa2016InvestigationIA}; and also Influenza as another influential airborne disease, both which also amplify the effects of Dengue and COVID-19 on the Brazilian population.

Studying those socioeconomic components as possible means of understanding the main factors of propagation could in turn increase the accurateness of predictions for those diseases, which would likely enable faster responses to new outbursts, as demonstrated in the case of malaria forecasting in Guyana \cite{Menkir2021Nov}. Previous studies on the spreading of dengue in Brazil shown that geographic proximity and hierarchical levels of influence between cities are impactful in the transmission process, with highly influential cities with many transport links having increased odds of an outbreak. \cite{Lee2021Dec}. Also, the readiness for a given country to respond and its vulnerability to the effects of epidemic diseases have also been quantified in terms of several socioeconomic indicators \cite{Chan2013Feb, Jain2019Dec}. For this study, the gross domestic product (GDP) of each municipality will be considered as the chosen indicator of economic growth.


In this work, the aim was to develop and apply a strategy utilizing the underlying information contained in socioeconomic and geographic data from Brazilian cities as a way to increase the effectiveness of predictions for diseases, while also verifying the impact of this protocol for intrinsically different conditions. Multiple decision tree frameworks were employed and analyzed under a criterion that includes the seasonal naïve baseline, selecting the best model for each city via cross-validation and then evaluating on a hold-out test set. Predictions for COVID-19 notably benefit from this methodology, while predictions for Dengue, Zika and Influenza benefit less from socioeconomic and geographic associations.

\section{Materials and methods}

\subsection{\label{sec:data} Data}

Weekly cases for the diseases were sourced by the official Brazilian government database. Dengue, Zika and Influenza registers can be found on the System of Information on Aggravations and Notifications (\textit{Sistema de Informação de Agravos de Notificação}, \textit{SINAN}) \cite{Datasus_site} and also on the official government panel for COVID-19 data \cite{Covid_weekly_data}. As for geographic and economic data, the information utilized to generate the correlations between cities can be found on IBGE platform (\textit{Instituto Brasileiro de Geografia e Estatística}) \cite{IBGE_site}, with data ranging from 2014 to 2020. The latitude, longitude, and GDP per capita of all available cities in this time frame were considered, with the geographic distance between municipalities calculated through the euclidean distances between latitude and longitude, while the yearly GDP data for each city was treated as time series and similarities were defined as discussed in Section \ref{sec:expansion}.

After combining the available cities on DataSUS with IBGE's database, and also accounting for all years in the time range considered for Dengue, Zika and Influenza, the analysis respectively accounted for 1804, 211 and 274 cities that fit this criteria, while the COVID-19 database encompass 5565 unique cities: notice that, although with much larger coverage on Brazilian cities, the endemic diseases have measurements on a greater array of years. Data was then split into a training set (Dengue: Jan. 2014 - May. 2020; Zika: Jan 2016 - May. 2022 ; COVID-19: Mar. 2020 - Feb. 2023; Influenza: Jan 2013 - Jul. 2018) and a hold-out test set (Dengue: Jun. 2020 - Dec. 2021; Zika: Jun. 2022 - Dec. 2023 ; COVID-19: Mar. 2023 - Dec. 2023; Influenza: Aug. 2018 - Dec. 2019). The data was normalized to have the maximum cases count of one for each trained model. 

For the four diseases included in this project, the skewness, mean and standard deviation of data was also included on Table \ref{tab:table_stats}.

\begin{table}[!t]
	\caption{\label{tab:table_stats}%
		Basic statistic description for the weekly number of cases to the investigated diseases.
	}
	\begin{tabular}{llll}
     \centering
		\textrm{Disease}&
		\textrm{Mean}&
		\textrm{Maximum}&
        \textrm{Skewness}\\
		\hline
		Dengue & $178 \pm 566$  & 13540 & $5.3 \pm 1.9$  \rule{0pt}{2.6ex}\\
		\hline
        Zika & $85 \pm 244$  & 2472 & $5.3 \pm 2.0$  \rule{0pt}{2.6ex}\\
		\hline
		COVID-19 & $552 \pm 2680$  & 107057 & $4.5 \pm 2.5$  \rule{0pt}{2.6ex}\\
        \hline
        Influenza & $18 \pm 51$  & 885 & $3.8 \pm 1.2$  \rule{0pt}{2.6ex}\\
	\end{tabular}
\end{table}

\subsection{\label{sec:model} Methods}



\begin{figure*}[!t]
        \centering
        {\includegraphics[width=1.5\columnwidth]{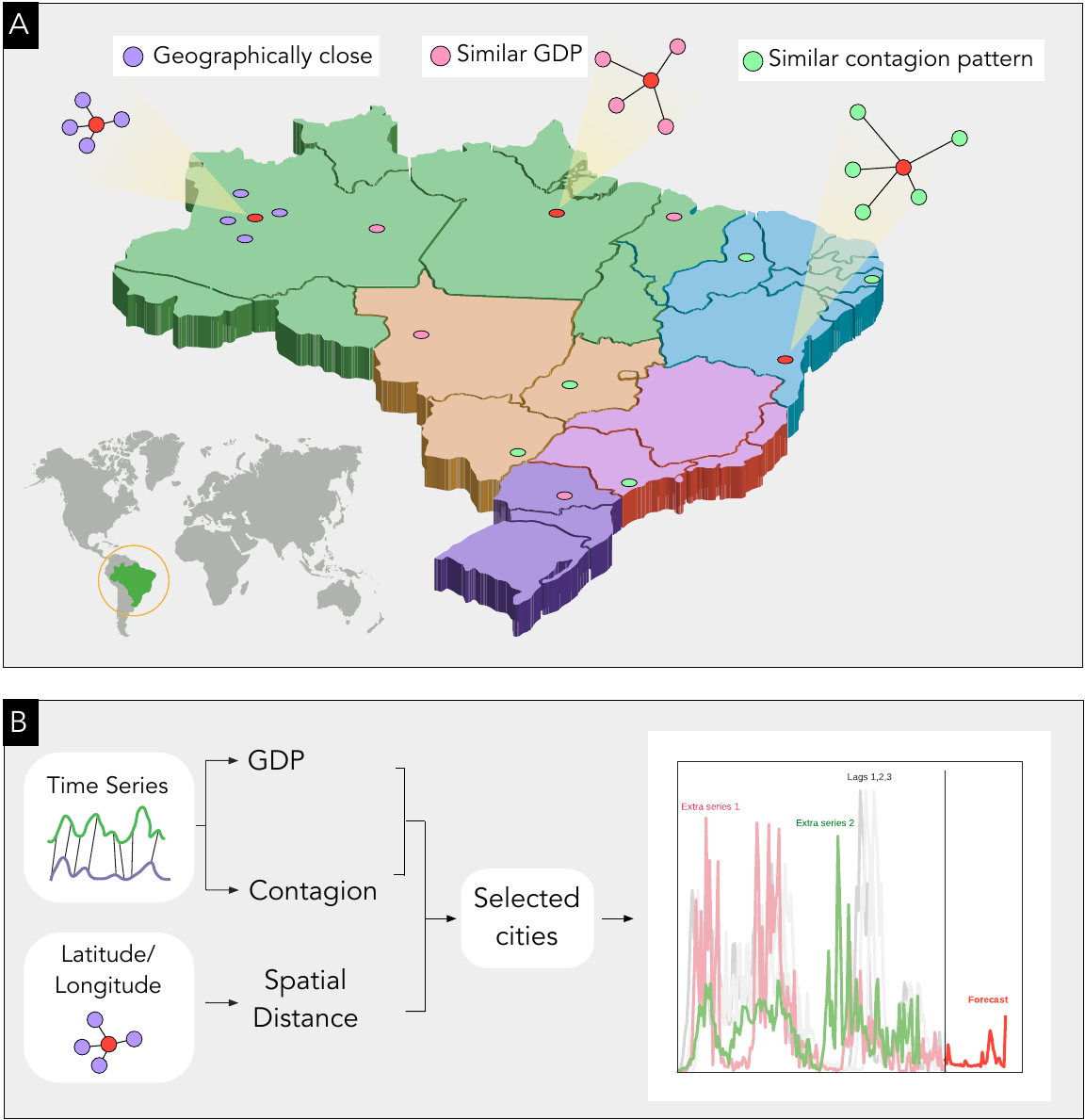} }
        \caption{Visual representation of the feature engineering method presented in Section \ref{sec:expansion}. (A) exemplifies how cities are selected according to spatial distance or similarities between time series representing the GDP or outbreak size evolution. (B) The prediction is performed by combining the time series of the selected cities. The forecasting on the target city is shown in red.}%
        \label{fig_scheme}%
\end{figure*}

We employed the predictive algorithms Random Forests and XGBoost for all predictive tasks, where the training set used 5 delays (or lags) for the disease's time series. These two models function as ensembles of decision trees with intrinsically different methods of achieving forecasts. In all of them, the trees are generated by selecting subsets of observations through sampling with replacement methods, in this work using a random subset of features in each splitting process and selecting the best fit using Mean Absolute Error (MAE).

Random Forests \cite{Breiman2001Oct} determine the final result using a combination of multiple decision trees, being an approach with few parameters to tune throughout the training processes and thus presenting a robustness to over-fitting. 
As for the Gradient Boosting Regression method XGBoost \cite{Chen2016Aug} chosen, the trees are built individually, where weights are added depending on the performance shown by that tree to a given example, that is, the ensemble will be able to account for higher variability in data by evaluating trees based on the difficulty of prediction for those examples.

Mean absolute scaled error (MASE) was used to evaluate the performance of predictions, due to its interpretability and scale-invariant properties \cite{Metricas_paper}.

	\begin{equation}
		\label{eq:MASE}
		MASE =  \frac{ \frac{1}{J} \sum_j |\hat{y}_j - y_j|}{ \frac{1}{T-m} \sum_{t=m+1} ^T |y_t - y_{t-m}|},
	\end{equation}
    \noindent
    where the numerator is just the mean average error (MAE).
    
MASE's interpretability comes from its direct link to the performance of the seasonal naïve model for a given city, which is defined in the denominator of Eq. \ref{eq:MASE}. Models with $MASE < 1$ will outperform the naïve forecast. Moreover, MASE is scale-invariant, a known limitation of MAE when comparing multiple time series with varying amplitudes, as was observed for all data and illustrated on Figure \ref{fig:series_examples}. It is also symmetric and robust when the predicted value $\hat{y}$ approaches 0, which is often the case in epidemiological records. The algorithm employed will take $m$ to be the same length of the prediction window for each disease.

In order to evaluate the performance of the implemented models, we began by defining the hyper-parameters of each model and using an exhaustive search method over those to define the optimal combination for the trained models, alongside employing cross-validation with four splits of the train data in each case. These parameters are listed in the following table.

\begin{table}[!t]
	\caption{\label{tab:table2}%
		Parameters for each regression model. Four splits of the train set were used for cross-validation.
	}
	\centering
	\begin{tabular}{lll}
		\textrm{Algorithm}&
		\textrm{Hyperparameter}&
		\textrm{Values}\\
		\hline
		RF & number of trees:    & [25,50,100,150,200] \rule{0pt}{2.6ex}\\
		& maximum tree depth: & [2,4,None] \\
		\hline
		XGBoost & number of trees:    & [25,50,100,150,200] \rule{0pt}{2.6ex}\\
		& maximum tree depth: & [2,4,None] \\
		& learning rate:      & [0.001, 0.005, 0.01]\\
	\end{tabular}
\end{table}

It is important to note that the cross-validation model in this approach selects the best hyperparameters for each city independently from the rest of the data set, both for the direct prediction baseline and for the methodology described in Sec. \ref{sec:expansion}.  

The data was filtered to exclude cities with anomalies, or outliers, defined here as cases outside the expected distribution in the training with z-score $z>4$ observed on the hold-out test set of each disease (see Sec. \ref{sec:data}). Despite all diseases displaying non-normal distribution of cases, according to the mean skewness on Table \ref{tab:table_stats}, the second moment of these distributions can still present an informed criteria according to Chebyshev's inequality, where at least $94\%$ of data would have $\sigma=4$ distance from the mean. As a deviation measurement, the implications to forecasting on cities with patterns that would not be seen by the models will be discussed in Sec. \ref{sec:discussion}. Table \ref{tab:best_model} and Fig. \ref{fig:results_all_models} describes the precision of the evaluated regression models for time series with and without anomalies present. 


\subsection{\label{sec:expansion} Feature Engineering}
The main objective of this study was to evaluate the predictive performance of different approaches for selecting features from related cities. We compared the baseline model described above to models that incorporate features from cities with correlated time series, represented in Figure \ref{fig_scheme} and defined below. 

\begin{enumerate}
	\item Geographic proximity between cities, with the euclidean distances being calculated from IBGE's data described in Sec. \ref{sec:data}. This allows one to generate a network of municipalities using only distances as the connection criteria, that is, the closest city to a given target will be part of its neighborhood;
 
	\item Optimal match distance calculated through the use of Dynamic Time Warping \cite{DTW_article} for both yearly GDP data for each city from the IBGE database (see Sec. \ref{sec:data});
 
    \item Same procedure as previously, now employed to evaluate the diseases' time series. This is done as a way to compare the patterns of contagions between cities, generating a non-informed baseline for the previous two selection criteria.
\end{enumerate}

To quantify the optimal match distances for diseases cases and GDP data, dynamic time warping is employed as a method to evaluate which of the time series will have the minimum traversal cost to a given target. Defining such traversals as $T = ( (i_1, j_1), ..., (i_t, j_t) )$, with $i_1 = j_1, i_t = n, j_t = m$ for each time step $k \in {1,...,t-1}$, the algorithm will then minimize the function:

	\begin{equation}
		\label{Eq:DTW_min}
		\sum_{k=1}^{t} d \left( p_{i_k} , q_{j_k} \right) ,
	\end{equation}
\noindent
where $p_{i_k}$ and $q_{j_k}$ are points in the curves $P = (p_1, ..., p_n) $ and $Q = (q_1, ..., q_m)$ that are being analyzed. The optimal match will be denoted as DTW$(P,Q)$.
Notice that Eq. \ref{Eq:DTW_min} requires a choice of distance metric $d(.,.)$, which in this work was taken to be the Euclidean metric, where $d(x,y) = || x-y ||_p$.

For all methods, new features made from the time series of cities that fulfilled the criteria above were added to the training sets, using up to the top three cities' time series with optimal, or minimal, distances with respect to a given target.


\section{Results}


To first generate a baseline, the selection of optimal hyperparameters for each individual city was performed with two regression models: Random Forests and XGBoost. The resulting best model on this validation stage was then evaluated on the test set slice of the city being studied. Table \ref{tab:best_model} show the average MASE performance of this approach for all cities.

\begin{table}[!t]
	\centering
	\caption{\label{tab:best_model} MASE performance on the train and hold-out test set of each regression framework for all diseases.}
	
		\begin{tabular}{c | c | c  c | c  c }
			\hline\hline
			\multicolumn{1}{c|}{} & \multicolumn{1}{c|}{} & \multicolumn{2}{c|}{$z < 4$} & \multicolumn{2}{c}{$z \geqslant 4$}\\
			
			\multicolumn{1}{c|}{Disease} & \multicolumn{1}{c|}{Algorithm} & \multicolumn{1}{c}{Train} & \multicolumn{1}{c|}{Test} &\multicolumn{1}{c}{Train} & \multicolumn{1}{c}{Test}\\[1ex]
			
			\hline
			Dengue	& \textbf{Random Forests} & 0.697 & \textbf{0.552} & 0.694 & 1.148 \rule{0pt}{2.6ex} \\
					& XGBoost 		          & 0.871 & 0.586 & 0.857 & 1.228 \\[1ex]
			  \hline
            Zika	& \textbf{Random Forests}  & 1.100 & \textbf{0.604} & 1.040 & 1.044 \rule{0pt}{2.6ex} \\
			 			& XGBoost		       & 1.287 & 0.703 & 1.199 & 1.134 \\ [1ex]
            \hline
			COVID-19	& \textbf{Random Forests}      & 0.605 & \textbf{0.362} & 0.605 & 0.423 \rule{0pt}{2.6ex} \\
			 			& XGBoost 	                   & 0.643 & 0.508 & 0.643 & 0.566 \\ [1ex]
            \hline
            Influenza	& Random Forests       & 0.798 & 1.070 & 0.786 & 1.274 \rule{0pt}{2.6ex} \\
			 			& \textbf{XGBoost} 	   & 0.866 & \textbf{1.061} & 0.847 & 1.277 \\ [1ex]
       
		\hline\hline	
		\end{tabular}
\end{table}

As shown in Table \ref{tab:best_model}, for Dengue and Zika the Random Forest algorithm performed best both in cities displaying anomalies or those without then in the test set; while for COVID-19 and Influenza data, XGBoost was the best overall model for cities. Nonetheless, excluding Zika, applying the models for the dataset that included anomalous cities significantly reduced the observed accuracy of the prediction. 

\begin{figure*}[t!]
	\centering 
	\begin{subfigure}[t]{0.48\textwidth}
		\centering
		\includegraphics[width=\textwidth]{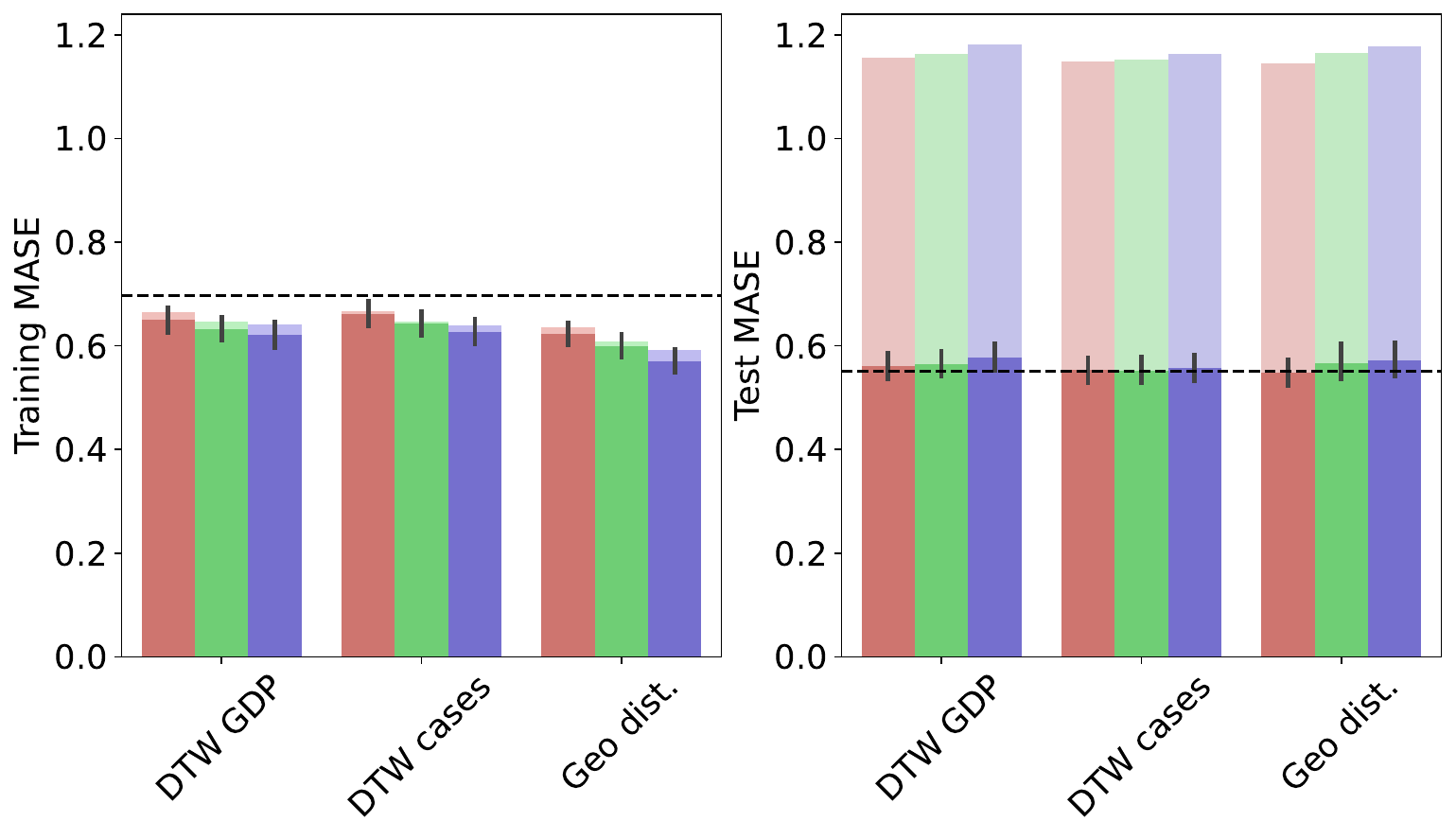}
		\subcaption{ Dengue forecasts using Random Forests.}
        \label{fig:res_dengue}
	\end{subfigure}
	\begin{subfigure}[t]{0.48\textwidth}
		\centering
		\includegraphics[width=\textwidth]{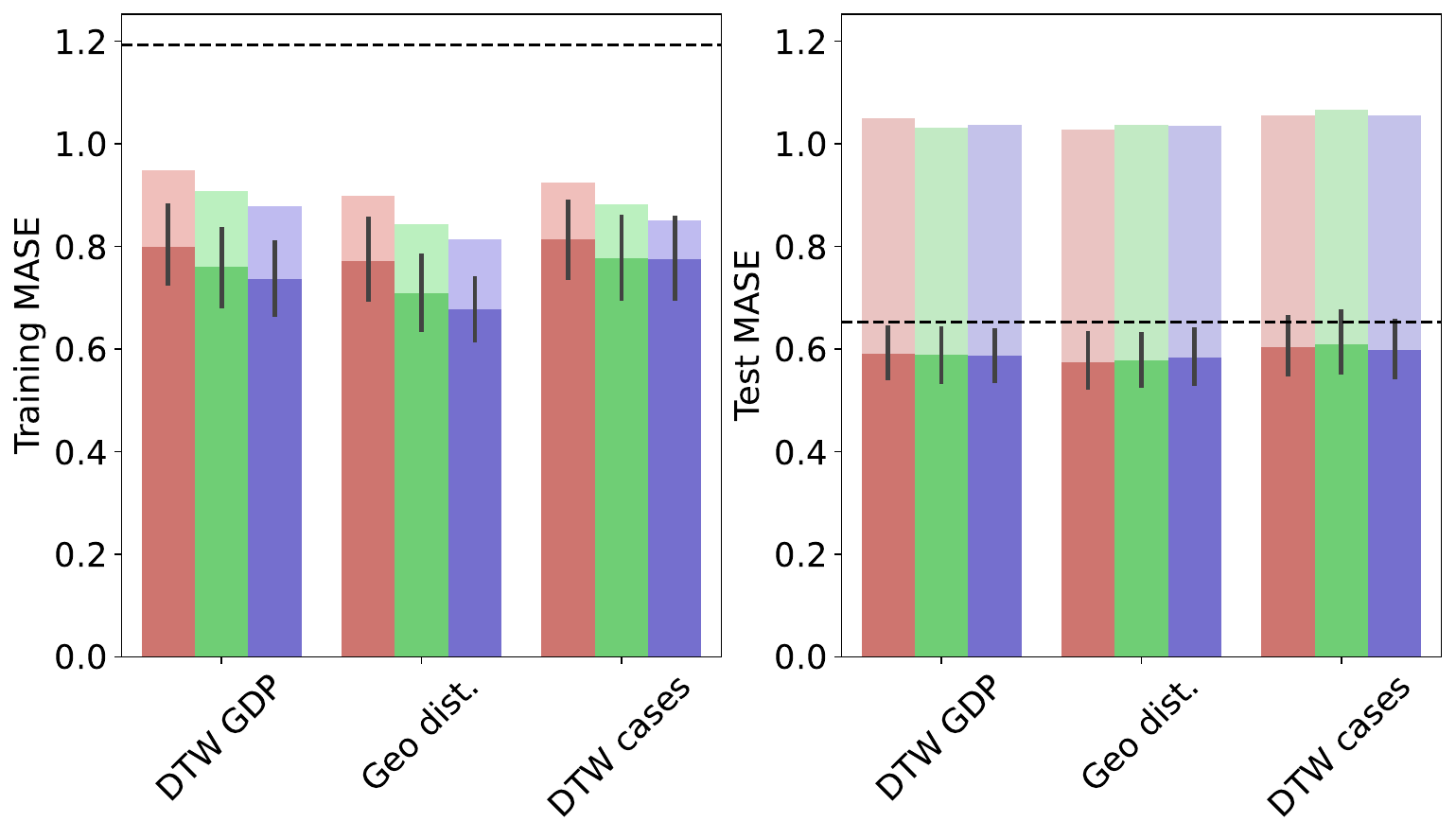}
		\subcaption{ Zika forecasts using Random Forests. }
        \label{fig:res_zika}
	\end{subfigure}
    \begin{subfigure}[t]{0.48\textwidth}
		\centering
		\includegraphics[width=\textwidth]{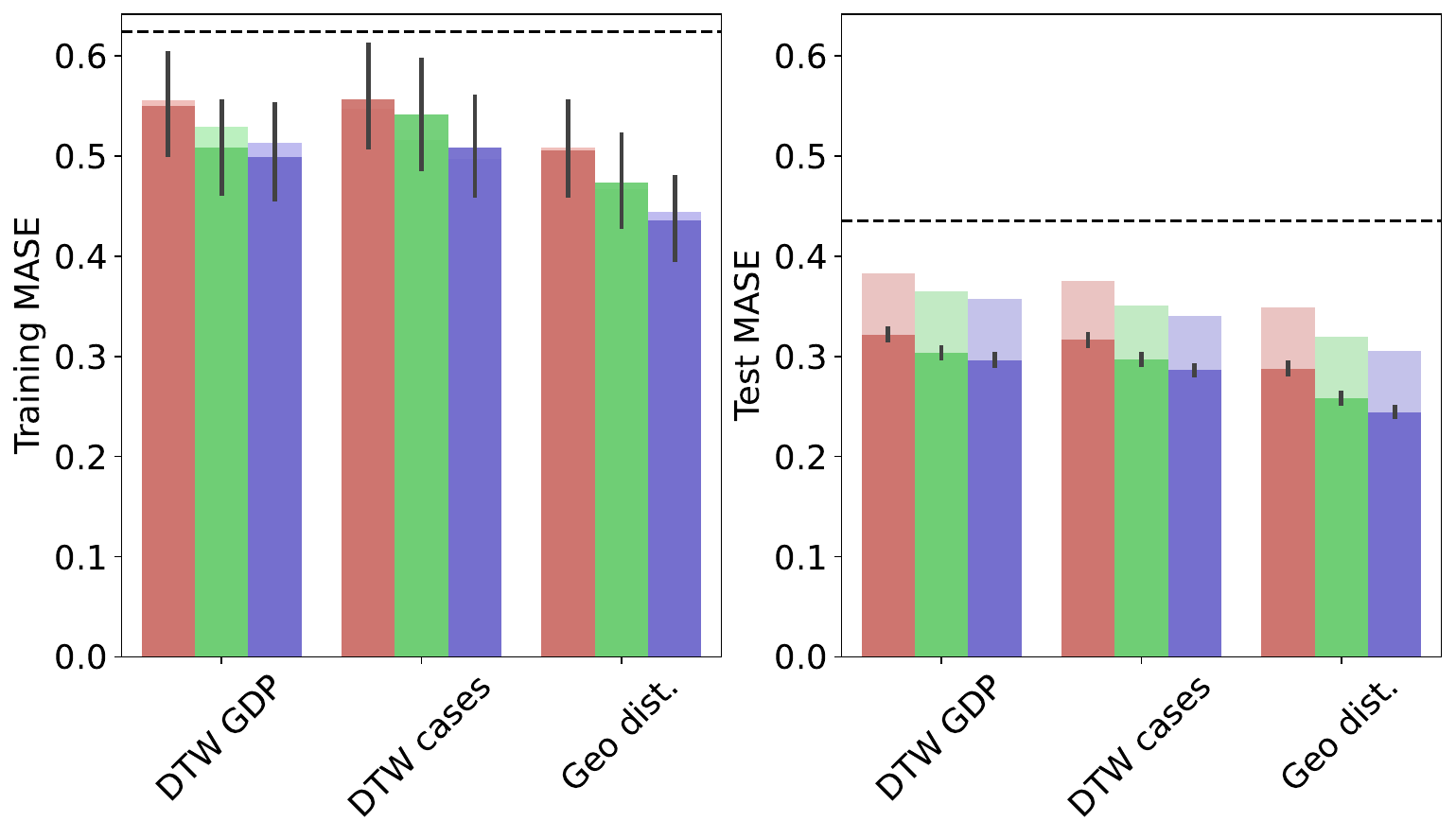}
		\subcaption{ COVID-19 forecasts using Random Forests.}
        \label{fig:res_covid}
	\end{subfigure}
    \begin{subfigure}[t]{0.48\textwidth}
		\centering
		\includegraphics[width=\textwidth]{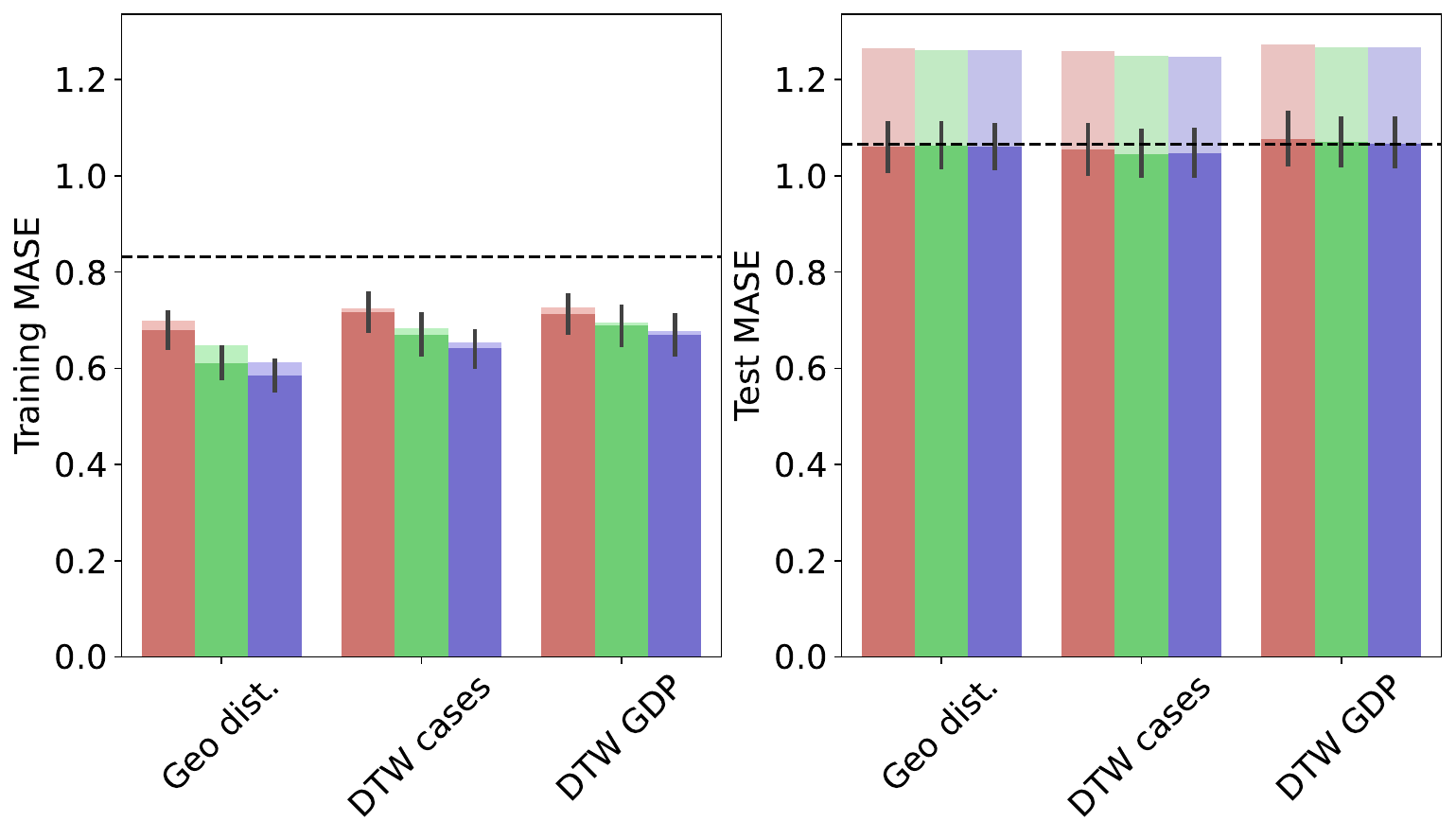}
		\subcaption{ Influenza (pre-COVID-19) forecasts using XGBoost. }
        \label{fig:res_flu}
	\end{subfigure}
	\caption{\label{fig:results_all_models} MASE performance using different numbers of features on the training set, where the quantity is represented by the colours of the bars (red: 1, green: 2, purple: 3), for all diseases studied. Solid colours indicate the use of the dataset without anomalous cities ($z<4$), while transparent bars show all data. Dashed lines represent the baseline model.}
\end{figure*}



The results shown in Figs. \ref{fig:results_all_models} used the best regression models from the baseline, respectively Random Forests for Dengue, Zika and COVID-19, and XGBoost for Influenza, along with the expanded train sets generated through the methodology described in Sec. \ref{sec:expansion} for the three different association criteria. The parameter optimization and cross-validation were executed independently from the baseline selection. 

When considering the augmentation of the training set with the proposed methodology, simulations performed for Dengue and Influenza cases do not display considerable benefits of increasing the features past the initial five lags from the target time series, given that all results are within the uncertainty range and are comparable to the baseline performance, shown in Figures \ref{fig:res_dengue}, \ref{fig:res_zika}, \ref{fig:res_flu}. 

COVID-19 and Zika predictions including such features shown in Fig. \ref{fig:res_covid}, on the other hand, notably increases the regression effectiveness, and more evidently for the hold-out test set for the pandemic, with geographical associations resulting in the best performance for both cases. Overall, aside from Influenza, all results without including anomalies in the test set displayed higher accuracy than the seasonal naïve model. 

\begin{figure*}
	\begin{subfigure}[t]{0.22\textwidth}
		\includegraphics[width=\textwidth]{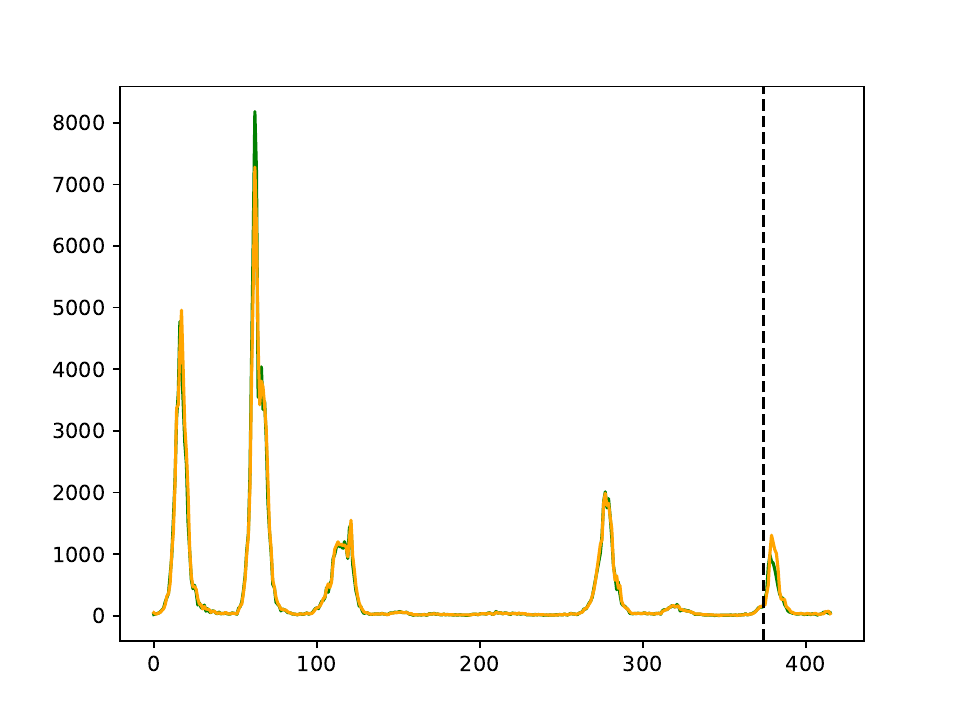}
		\caption{Dengue on São Paulo, SP. MAE: 64.31, MASE: 0.63.}
	\end{subfigure}
    \begin{subfigure}[t]{0.22\textwidth}
		\includegraphics[width=\textwidth]{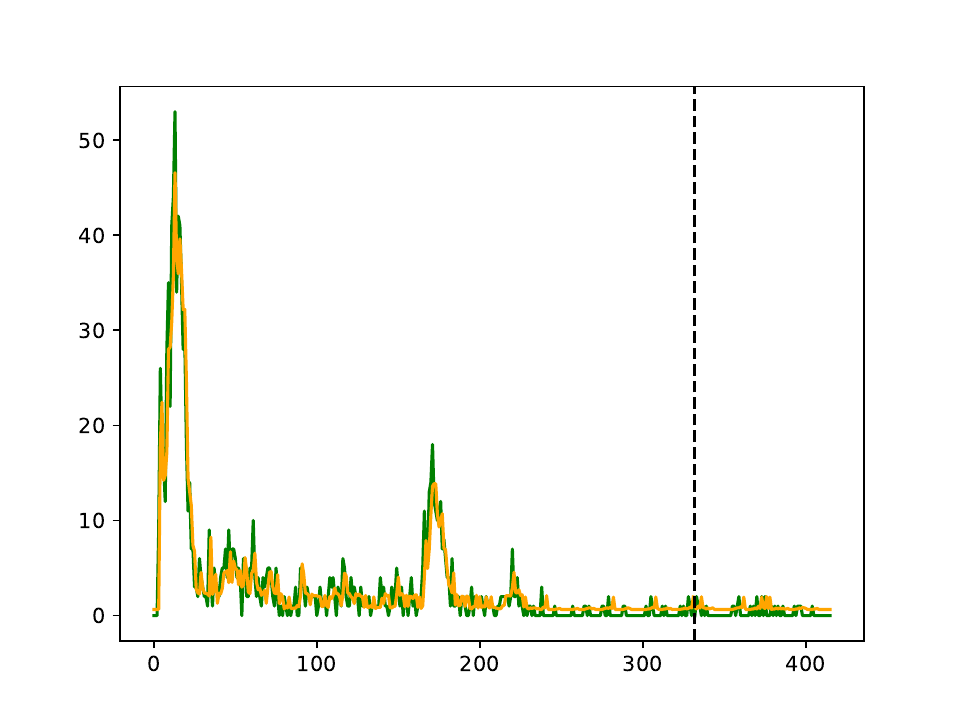}
		\caption{Zika on São Paulo, SP. MAE: 0.70, MASE: 0.41.}
	\end{subfigure}
	\begin{subfigure}[t]{0.22\textwidth}
		\includegraphics[width=\textwidth]{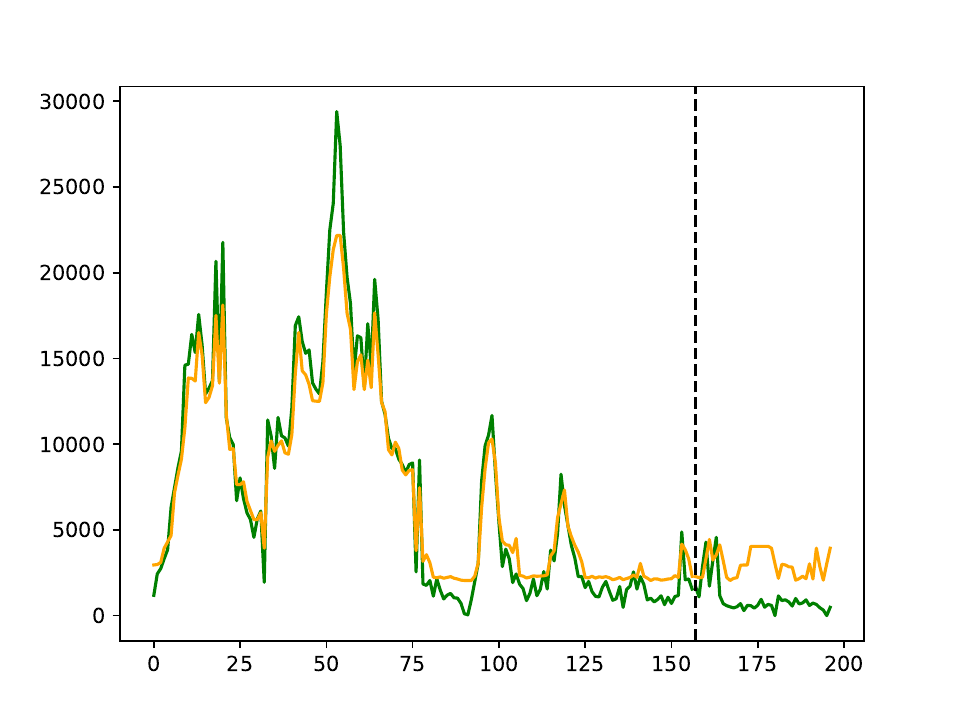}
		\caption{COVID-19 on São Paulo, SP. MAE: 2146.37, MASE: 1.33.}
	\end{subfigure}
    \begin{subfigure}[t]{0.22\textwidth}
		\includegraphics[width=\textwidth]{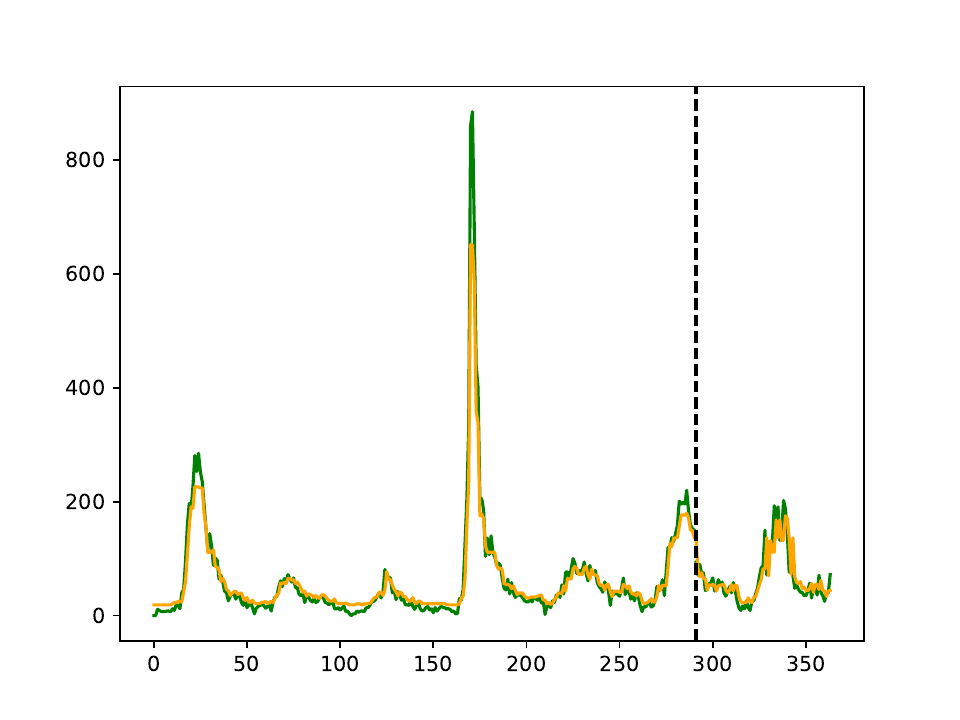}
		\caption{Influenza on São Paulo, SP. MAE: 18.94, MASE: 1.22.}
	\end{subfigure}
	
	\begin{subfigure}[t]{0.22\textwidth}
		\includegraphics[width=\textwidth]{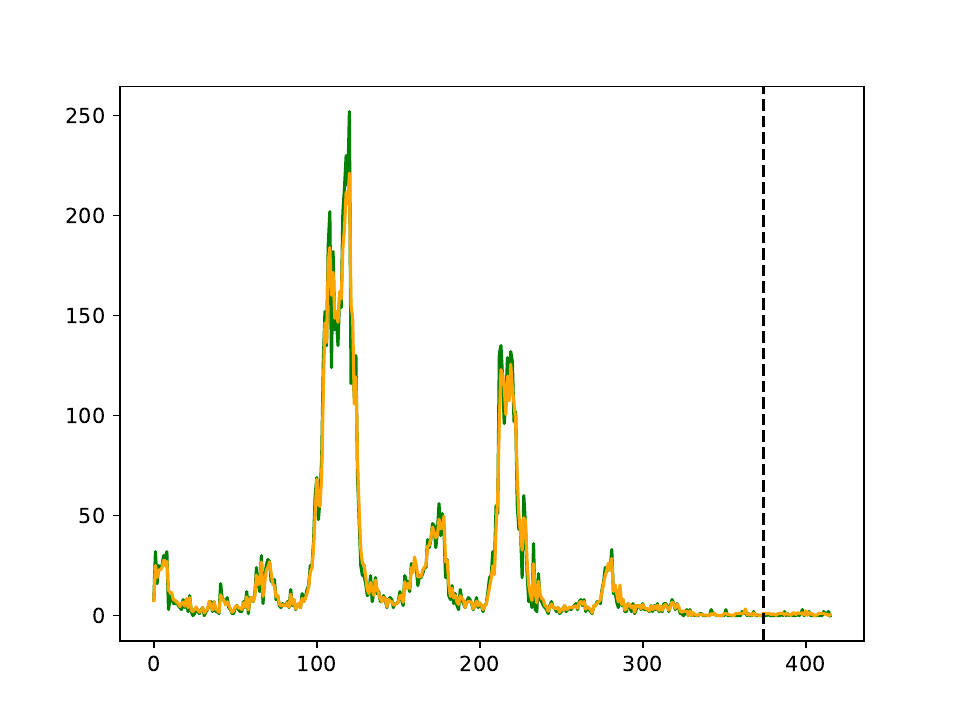}
		\caption{Dengue on Niterói, RJ. MAE: 0.76, MASE: 0.11.}
	\end{subfigure}
    \begin{subfigure}[t]{0.22\textwidth}
		\includegraphics[width=\textwidth]{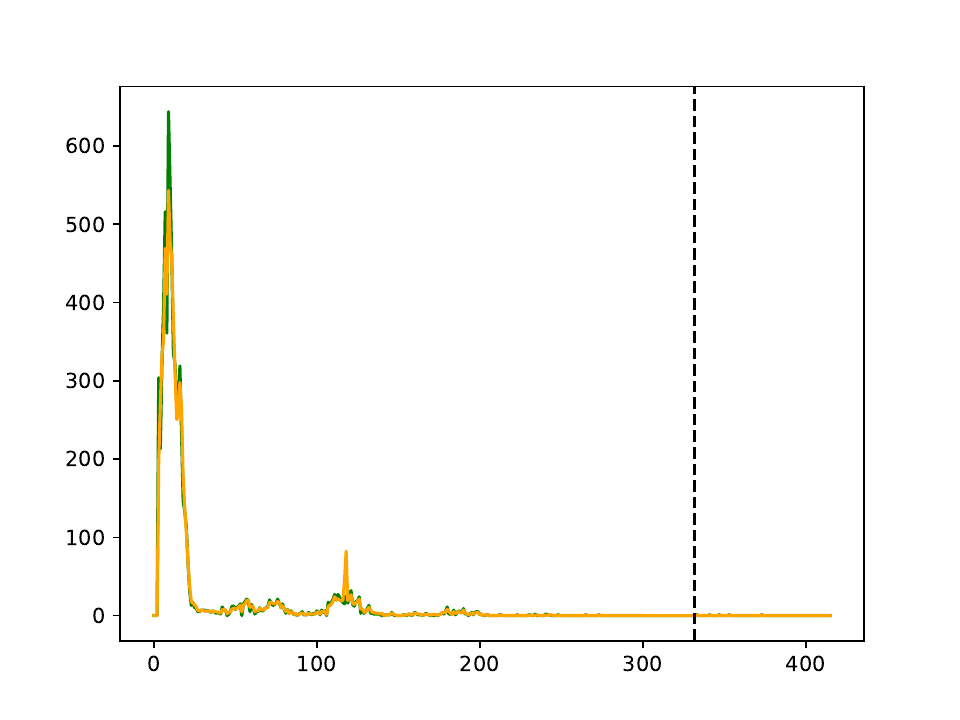}
		\caption{Zika on Niterói, RJ. MAE: 0.16, MASE: 0.02.}
	\end{subfigure}
	\begin{subfigure}[t]{0.22\textwidth}
		\includegraphics[width=\textwidth]{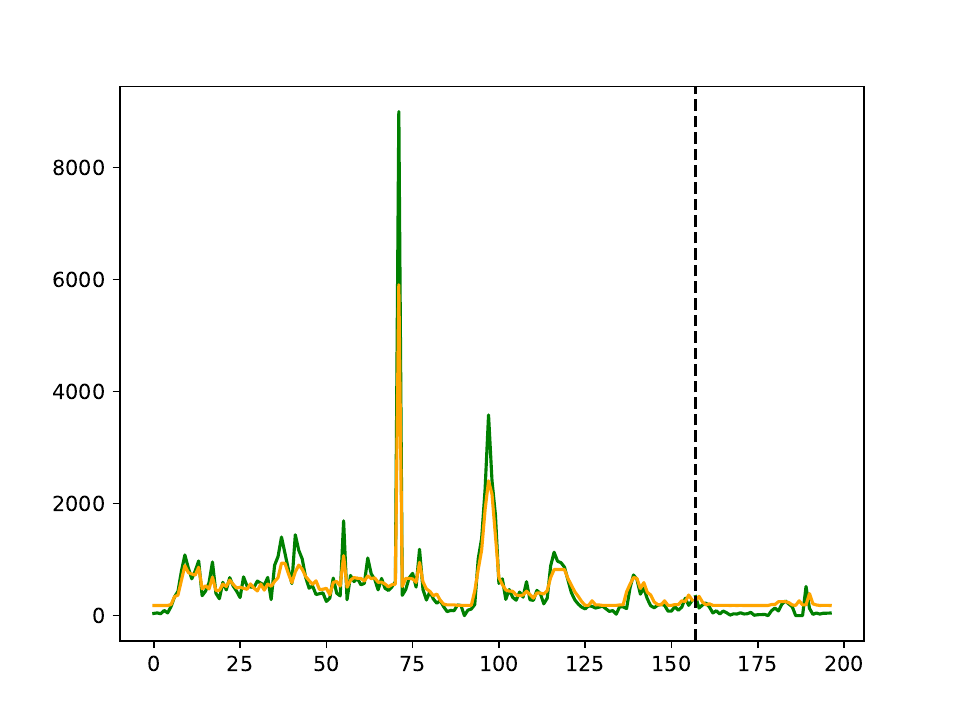}
		\caption{COVID-19 on Niterói, RJ. MAE: 137.34, MASE: 0.43.}
	\end{subfigure}
    \begin{subfigure}[t]{0.22\textwidth}
		\includegraphics[width=\textwidth]{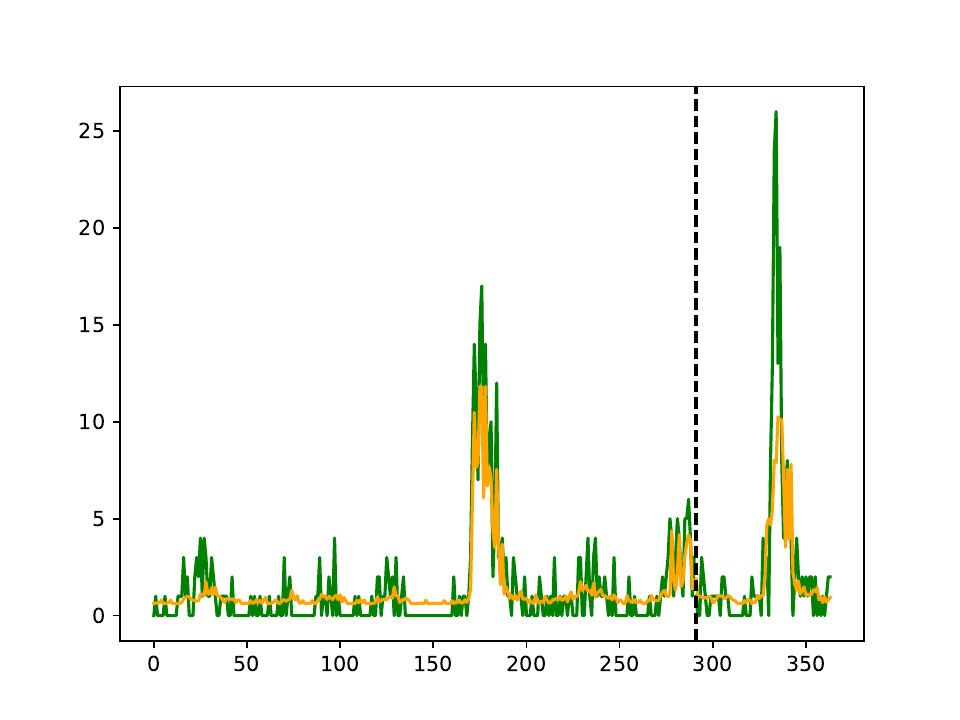}
		\caption{Influenza on Niterói, RJ. MAE: 1.66, MASE: 1.55.}
	\end{subfigure}
	
	\begin{subfigure}[t]{0.22\textwidth}
		\includegraphics[width=\textwidth]{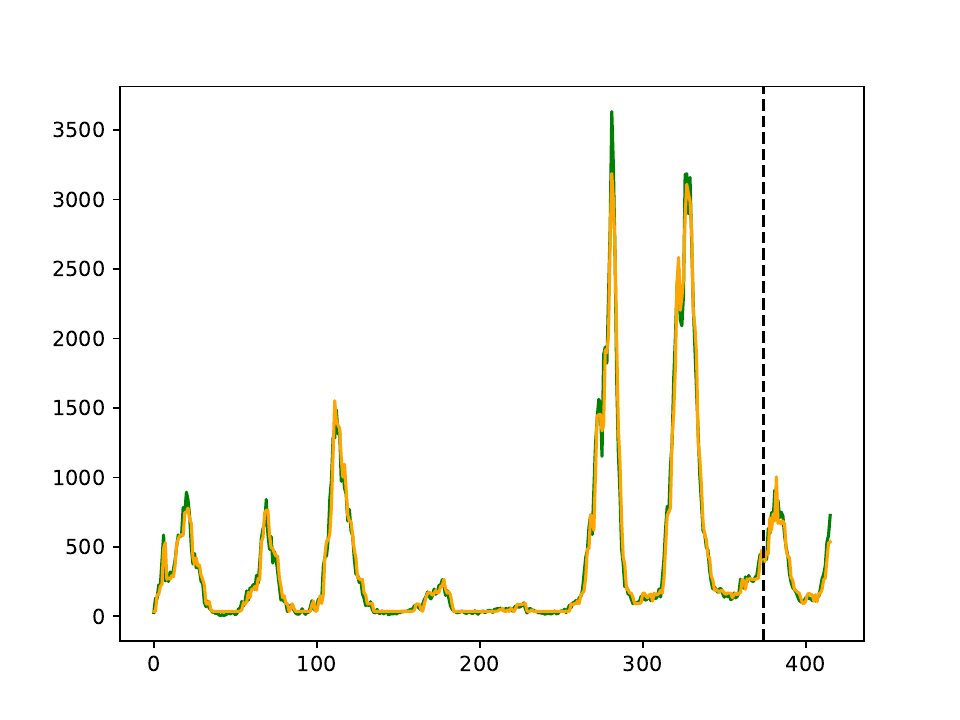}
		\caption{Dengue on Brasília, SP. MAE: 67.2, MASE: 0.89.}
	\end{subfigure}
    \begin{subfigure}[t]{0.22\textwidth}
		\includegraphics[width=\textwidth]{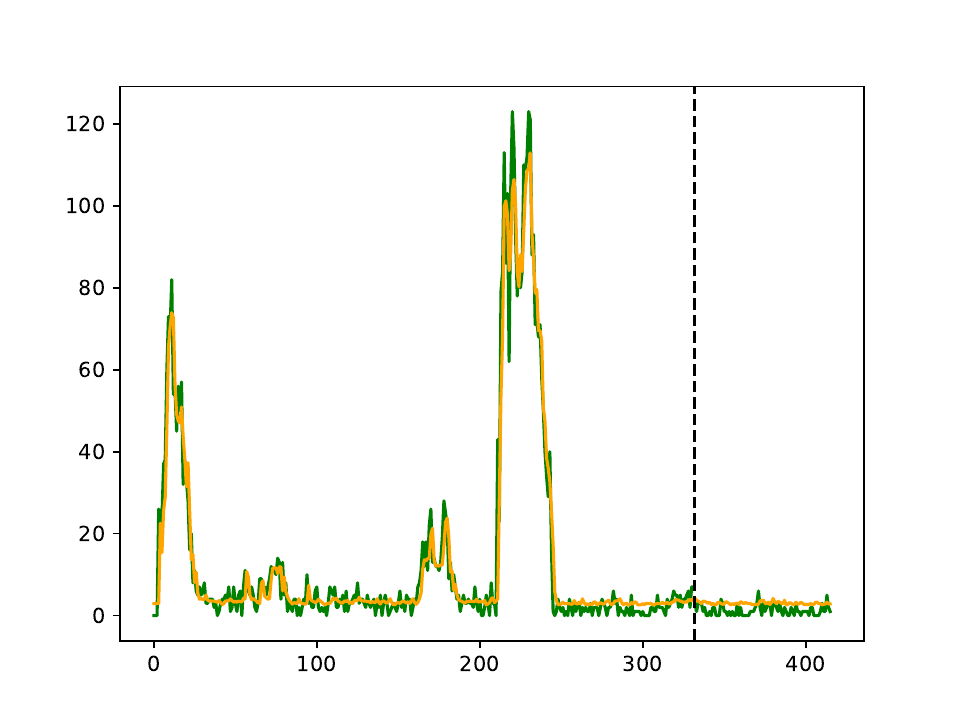}
		\caption{Zika on Brasília, SP. MAE: 1.90, MASE: 0.43.}
	\end{subfigure}
	\begin{subfigure}[t]{0.22\textwidth}
		\includegraphics[width=\textwidth]{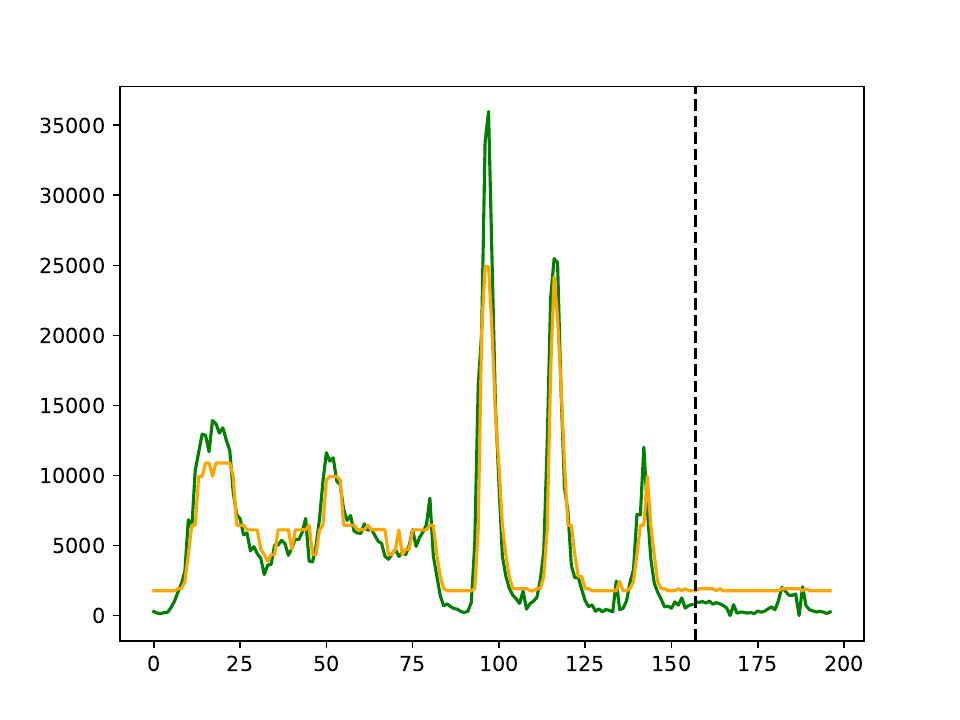}
		\caption{COVID-19 on Brasília, SP. MAE: 1164.87, MASE: 0.80.}
	\end{subfigure}
    \begin{subfigure}[t]{0.22\textwidth}
		\includegraphics[width=\textwidth]{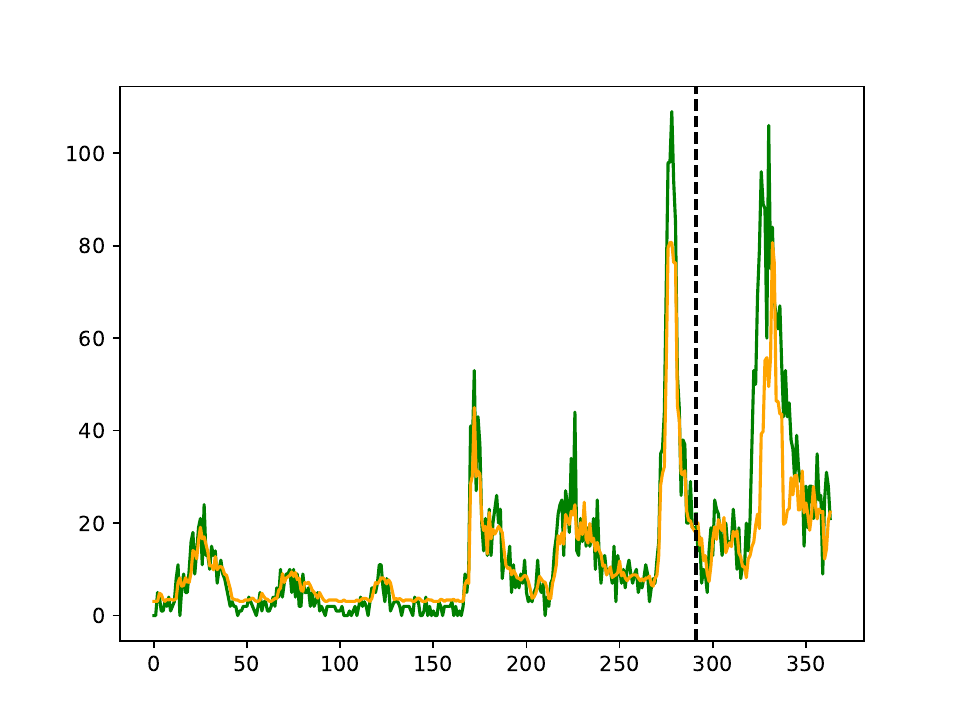}
		\caption{Influenza on Brasília, SP. MAE: 11.98, MASE: 2.84.}
	\end{subfigure}
    \caption{\label{fig:series_examples} Example predictions for three sample cities for all diseases, using the geographic distances as the aggregation method on the train set and including three new features from associated cities. From left to right: Dengue, Zika, COVID-19 and Influenza cases. None of these time series contain anomalous patterns in the hold-out test set. MAE and MASE results are presented for the test set.}

\end{figure*}

Fig. \ref{fig:series_examples} illustrates the forecasts, and also the mentioned advantages of using MASE as a evaluation metric instead of MAE, as discussed in Section \ref{sec:model}, for selected municipalities using the best models for each disease according to the results of Fig. \ref{fig:results_all_models}.

\section{Discussion}
\label{sec:discussion}


The main proposal of this method is to verify the potential of including variables that are known to be linked to a given disease's spread dynamic, thus not only creating predictions that are robust to reporting fluctuations, but also identifying ``signaling cities" and help understand the wave-like patterns of disease transmission across Brazil. The positive outcome of including disease's data from geographically close cities for Zika and COVID-19 can be interpreted as a reflection of such patterns, indicating that regional-level health policies also can be effective in the containment of those outbreaks.

It is important to also note the limitations of this study. The official reports on dengue, for example, could be instead a mislabel of Zika or Chikungunya, that shares some symptoms with Dengue, as shown by \cite{Pessa2016InvestigationIA}. Furthermore, the predictions for Dengue and Influenza were not substantially enhanced through this method, which could imply that the variables selected are less impactful for these endemic diseases, or the trends included in the correlated time series does not contain information that can positively influence the model. 

As for the use of GDP as the selection criteria for the training set, none of the considered diseases benefited significantly from including this metric. This implies that 
other, more specific measurements of social and economic aspects should be used instead. Chan et al. \cite{Chan2013Feb} demonstrated the effectiveness of using health and infrastructure indicators to the prediction of outbreaks in multiple countries, including Brazil.

Moreover, most common regression models, including decision trees, learn from seasonal patterns and trends, and as such would not be able to predict significant deviations from the training data should such pattern emerge in future measurements, as demonstrated by the notable difference in evaluating the performance on target series within or without the z-score threshold.

\section{Conclusions}
\label{sec:conclusion}
In this work, the predictive performance of machine learning models that incorporate information from related cities was evaluated by comparing three methods for selecting related cities, through similarity in geography, GDP and seasonal patterns in the data. We implemented these methods for four different diseases in Brazil. COVID-19 and Zika predictions improved when enriching the training data with features from geographically proximate cities, while dengue and influenza forecasts did not benefit significantly from the same procedure. Moreover, forecasting is improved when applying the models for data that does not include unseen variations on the test set, with better performance than baseline in those cases. These results suggest that predictive models incorporating information from related cities can help infectious disease forecasts and create more robust early warning systems for public health departments. 


Expansions of this work could be done in multiple ways. First, data describing the travel flux between cities could help clarify the association of distances with the spreading of diseases, and consequently the impact observed of the use of this property in the prediction of COVID-19. Moreover, other indicators could be applied to this methodology, such as the Gini coefficient, the existence and investment levels in public health measures and sanitary services along with their coverage in a given city, hospitalization and mortality rates of the disease under study, and also other structural indicators, such as communications networks. It also could be useful to consider climate variables, specially for diseases transmitted through non-human vectors, as done in \cite{Stolerman2019Aug}.

An alternative method that could be used to reduce the influence of outliers in the modeling process would be to use a \textit{cellwise robust} filtering task, where it would flag cells in the data matrix as outlying and down-weight the influence of outliers, such as proposed on \cite{Alqallaf2009Feb}. In the case of large datasets complying to sparsity requirements, a recent work could also provide further insights to the modeling process \cite{Bottmer2022Mar}.

Furthermore, the explainability of a acute outbreak would require causal inference of multiple factors that goes beyond the scope of this project; while a halt, or a markedly decrease, of cases may be related to the implementation of lock-downs and other containment measures and should be taken into account when asserting the accuracy of predictions. In this context, the proposed methodology will then be effective in scenarios where the epidemic does not include such variations, unless those changes in the disease's pattern can be explained by known data. These models will then provide useful insights into the diseases dynamics by employing variables known to be linked to it that also improve forecastings, which in turn could contribute as an additional information source for public health decision-making.


\begin{acknowledgments}
    Luiza Lober thanks the support given by São Paulo Research Foundation (FAPESP) (grants number 2022/16065-3 and 2013/07375-0). Francisco A. Rodrigues acknowledges CNPq (grant 308162/2023-4) and FAPESP (grants 20/09835-1 and 13/07375-0) for the financial support given for this research. This project was conducted with the computational resources of the Center for Research in Mathematical Sciences Applied to Industry (CeMEAI) funded by FAPESP, Grant 2013/07375-0. We would also like to thank Paulo Cesar Ventura for the thoughtful discussions and suggestions in the final stages of this work. 
\end{acknowledgments}

\section*{Code availability}
All data and code used in this work are publicly available at \url{https://github.com/luizalober/epidemics-using-features}.


\bibliography{dengue_network}

\end{document}